\documentclass{aa}

\usepackage{graphicx}
\usepackage{lscape}
\usepackage{txfonts}
\usepackage{natbib}
\bibpunct{(}{)}{;}{a}{}{,}

\begin{document}
   \title{Complexity of magnetic fields on red dwarfs\thanks{The spectra of the  stars  analyzed in this work are also available in electronic form
at the CDS via anonymous ftp to cdsarc.u-strasbg.fr (130.79.128.5)
or via http://cdsweb.u-strasbg.fr/cgi-bin/qcat?J/A+A/.}}
 
    \titlerunning{Complexity of magnetic fields on red dwarfs}  
    
    \author{N.~Afram\inst{1}
          \and S.~V.~Berdyugina\inst{1,2}
          }

   \institute{Leibniz-Institut f\"ur Sonnenphysik (KIS), Freiburg, Germany
    \and Kavli Institute for Theoretical Physics, UC Santa Barbara, CA, USA
         }
   \date{Received date; accepted date}

\abstract
{Magnetic fields in cool stars can be investigated by measuring Zeeman line broadening and polarization in atomic and molecular lines. Similar to the Sun, these fields are complex and height-dependent. Many molecular lines dominating M-dwarf spectra (e.g., FeH, CaH, MgH, and TiO) are temperature- and Zeeman- sensitive and form at different atmospheric heights, which makes them excellent probes of magnetic fields on M dwarfs.}
{Our goal is to analyze the complexity of magnetic fields in M dwarfs. We investigate how magnetic fields vary with the stellar temperature  and how "surface" inhomogeneities are distributed in height -- the dimension that is usually neglected in stellar magnetic studies.} 
{We have determined effective temperatures of the photosphere and of magnetic features, magnetic field strengths and filling factors for nine M dwarfs (M1--M7). Our $\chi^2$ analysis is based on a comparison of observed and synthetic intensity and circular polarization profiles. Stokes profiles were calculated by solving polarized radiative transfer equations.}
{Properties of magnetic structures depend on the analyzed atomic or molecular species and their formation heights. Two types of magnetic features similar to those on the Sun have been found: a cooler (starspots) and a hotter (network) one. The magnetic field strength in both starspots and network is within 3 kG to 6 kG, on average it is 5 kG. These fields occupy a large fraction of M dwarf atmospheres at all heights, up to 100\%. The plasma $\beta$ is less than one, implying highly magnetized stars.}
{A combination of molecular and atomic species and a simultaneous analysis of intensity and circular polarization spectra have allowed us to better decipher the complexity of magnetic fields on M dwarfs, including their dependence on the atmospheric height. This work provides an opportunity to investigate a larger sample of M dwarfs and L-type brown dwarfs.}

  
   \maketitle

\section{Introduction}   
Magnetic fields are pervasive in a variety of astrophysical objects, ranging from planets to stars, and to galaxies. They strongly influence the object's structure, dynamics, and evolution. This fact is now being incorporated into evolutionary models. Furthermore, increasing attention has been put on the impact of stellar activity on planets, where the host star's magnetic fields perturb the planetary magnetosphere, which is thought to be essential in providing a protective shield for the evolution of life.

One of the fundamental astrophysical challenges is the understanding of the generation, amplification, and complexity of stellar magnetic fields. Stellar surface activity arises in the presence of a convection zone, which appears in F stars and takes over the whole star roughly at a spectral type M3 to M4  \citep{chabrierbaraffe1997, westetal2008}. These fully convective low-mass main-sequence stars lack an interface layer at the bottom of the convection zone like in the Sun, where at least the cyclic part of the solar dynamo is believed to operate. Thus, the existence of activity on those stars requires an alternative mechanism to generate magnetic fields. It is still unknown how the dynamo mechanism on late-type dwarfs changes as the mass reduces from solar type to fully convective objects. For early-type M dwarfs, it is possible that both the solar type and turbulent dynamos coexist, whereas the fully convective late-type M dwarfs and brown dwarfs can certainly maintain only a turbulent (distributed) dynamo. 

M dwarfs are the most common type of stars in the galaxy. They are in the center of attention in the exoplanetary research field as they are numerous and can host Earth-size planets in the star's habitable zone. These can be detected even for low-luminosity M dwarfs \citep[e.g.,][]{angladaescudeetal2016, gillonetal2017}. Their strong activity also affects the planetary atmospheres and magnetospheres. In addition, the knowledge of the star's magnetic field complexity and topology, and possible changes thereof, can help to reduce the uncertainties in searching for exoplanets around M dwarfs.
  
The direct detection of stellar magnetism is still a difficult task. Even for our closest neighbor, the Sun, mysteries about the magnetic fields remain unsolved. Considering the amount of information on solar magnetic fields, it is evident that the gaps in our knowledge about magnetic fields on  M dwarfs are enormous. An overview of magnetic fields on cool stars can be found in \citet{berdyugina2005,berdyugina2009}, \citet{strassmeier2009,strassmeier2011}, and \citet{reiners2012}.

Stellar magnetic fields are estimated by using the Zeeman effect in atomic and molecular spectral lines. To fully describe the magnetic field vector, one needs to measure all states of line polarization described by the Stokes parameters I (total intensity or flux), Q \&\ U (linear polarization), and V (circular polarization). In solar observations, all Stokes parameters are measured quasi-simultaneously, but such a sequence of measurements is significantly longer in stellar observations due to flux limitations. Also, since linear polarization caused by magnetic fields is typically smaller than circular polarization, measuring Stokes Q \&\ U is often neglected. 

Due to these challenges and a lack of high spectral resolution stellar spectropolarimeters, many previous magnetic field studies were based on an analysis of Stokes I only until recently. The basis of such an analysis is a comparison of Zeeman broadened lines with large and small Land\'e factors, that is, magnetically sensitive and insensitive line profiles in observed spectra \citep{robinson1980}. A similar technique was applied  by \citet{basrietal1992} who analyzed the change in line equivalent widths to measure magnetic fields on cool stars. \citet{reinersbasri2006} applied this approach to M dwarf spectra of atoms and molecules.

Detecting Stokes V in stellar spectra is considered to be an unambiguous detection of a magnetic field, in the absence of possible systematic errors. To increase the detection sensitivity, various techniques averaging Stokes profiles of thousands of atomic lines were developed \citep[e.g.,~][]{donatietal1997,sennhauseretal2009}. The disadvantage of such global filtering is that the spectral information is lost and only an average atmosphere can be diagnosed. \citet{semel1989} and \citet{semeletal1993} studied Stokes V atomic spectra and proposed to deconvolve their time-series into surface magnetic field maps (Zeeman-Doppler Imaging, ZDI). This technique has been extensively employed in recent years \citep[e.g.,~][]{donatilandstreet2009, donati2013, morinetal2010}. 

Using only Stokes I or only Stokes V as well as average atomic spectral lines for measuring and mapping stellar magnetic fields have large limitations \citep[see, e.g.,][]{berdyugina2009}. In particular, these approaches do not fully describe magnetic fields in starspots, whose internal physical parameters remain unaccessible. 

While the atomic Zeeman effect can probe warmer regions on A to mid-M type stars, the use of molecular lines is of advantage  for studying cool objects/starspots in the optical and infrared wavelength regions, where atomic lines diminish their utility among the forest of molecular features. The power of molecular spectroscopy and spectropolarimetry  for studying stellar magnetic fields has been demonstrated during the last two decades. Thorough theoretical and observational studies of magnetic properties of diatomic molecules/radicals observed in spectra of the Sun and cool stars they have been established  as unique probes of physical parameters prevalent in the quiet Sun, sunspots and starspots, including temperature and magnetic field strength \citep{berdyuginaetal2000, berdyuginaetal2003, berdyugina2011}. The sensitivity of different molecules (such as MgH, TiO, FeH, and CaH) for probing magnetic fields in starspots on F, G, K, and M stars was recently demonstrated by \citet{aframberdyugina2015}. These developments have led to the first direct detections of magnetic fields in starspots on M dwarfs \citep{berdyuginaetal2006b} and on a brown dwarf \citep{berdyuginaetal2017,kuzmychovetal2017}. Theoretical developments and astrophysical applications have encouraged studies of magnetic properties of diatomic molecules in physics laboratory experiments, such as TiO \citep{virgoetal2005},  CrH \citep{chenetal2007}, MgH \citep{zhangsteimle2014}, etc. 
 
In this paper, we analyze Stokes I and Stokes V together, and a number of atomic and molecular lines simultaneously. Our goal is to shed light on the complexity of the magnetic fields observed on M dwarfs.  Here, when we speak of the complexity, we particularly mean how magnetic fields (their strengths and filling factors) are distributed on the surface (local horizontal plane) and within different layers of the stellar atmosphere (local vertical direction). This 3D structure can be unraveled by using molecular and atomic species with different formation depths and temperature sensitivities \citep{berdyugina2011}. 
 
To achieve this goal, we determined the effective (photosphere and spot) temperature, the magnetic field strength, and its filling factors for a sample of M dwarfs with spectral types ranging from M1 to M7. Following \citet{aframetal2008}, we carried out a $\chi^2$ minimization of discrepancies between observed and modeled intensity and circular polarization spectra featuring seven different molecular and atomic bands or lines that are magnetically sensitive. The sensitivity of the considered molecules for probing magnetic fields in starspots was studied by \citet{aframberdyugina2015}. The right choice of magnetically sensitive lines is very important for the magnetic field determination because the inclusion of many marginally magnetically sensitive lines in any analysis only dilutes the result.

\begin{table*} 
\caption{Red dwarf data from the online archive of the Canadian Astronomy Data Centre. Some of the data used in this paper were previously published by \citet{berdyuginaetal2006b,morinetal2010,fouqueetal2018}. }
\smallskip
\label{tab:tabdata}
\centering
\begin{tabular}{lccccccl}
\\
\hline
\noalign{\smallskip}
Name &Spectral Type &v$\sin i[km/s]$&S/N (at 800nm)& Telescope/Instrument &  Obs. Date UT &  P.I. Name \\
 \hline
 \noalign{\smallskip}
AU Mic         & M1  & 7 & $\sim$920  &CFHT/ESPaDOnS & 2005-07-16  & Berdyugina \\
GJ 360         & M2  & 5 & $\sim$380  &CFHT/ESPaDOnS & 2014-11-11  & Malo \\
AD Leo         & M3.5& 3 & $\sim$770  &CFHT/ESPaDOnS & 2007-06-25  & Harrington \\
EV Lac (Gl 873)& M3.5& 5 & $\sim$800  &CFHT/ESPaDOnS & 2005-07-17  & Berdyugina  \\
YZ Cmi         & M4.5& 7 & $\sim$330  &CFHT/ESPaDOnS & 2006-02-07  & Forveille  \\
GJ 1224        & M4.5& 0 & $\sim$200   &CFHT/ESPaDOnS & 2008-06-25  & Morin  \\
GJ 1245B       & M5.5& 7 & $\sim$180 &CFHT/ESPaDOnS & 2007-10-03  & Wade  \\
CN Leo (Gl 406)& M5.5& 3 & $\sim$210  &CFHT/ESPaDOnS & 2008-03-22  & Morin  \\
VB 8           & M7.0& 5 & $\sim$100  &CFHT/ESPaDOnS & 2009-05-04  & Morin  \\
%
\hline
\end{tabular}
\end{table*}

The paper is organized as follows: In Sect.~\ref{data}, a description of the observational data is given, and in  Sect.~\ref{moldat}, the employed molecular and atomic data for FeH, CaH, MgH, TiO, Fe~I, and Ti~I lines are described. Section~\ref{method} presents our method: we have modeled the observed line profiles employing our numerical codes based on the full Stokes parameter radiative transfer in atomic and molecular lines in the presence of magnetic fields \citep{berdyuginaetal2003}. In Sect.\ref{procedure}, we explain each step of our procedure. An analysis of sensitivities of our results is given in Sect.~\ref{error}. We discuss our results for each parameter separately in Sect.~\ref{red}, and summarize our conclusions in Sect.~\ref{conc}.

\section{Observational data}\label{data}

The data used throughout this paper are described in Table~\ref{tab:tabdata}  and contain a sample of relatively slowly rotating early- to late-M dwarfs. The calibrated data are taken from the online archive of observations with the spectropolarimeter ESPaDOnS \citep{donati2003} at the 3.6~m Canada-France-Hawaii Telescope (CFHT). The resolving power for CFHT/ESPaDOnS is approximately 68,000,  and the observed wavelength region covers the range from 3700 to 10480 \AA, including many molecular (FeH, CaH, MgH, and TiO) and atomic (Fe and Ti) lines.

Stars were chosen to evenly cover the spectral type range from M1 to M7. Properties for the selected targets, such as the ages and activity levels, can be found in the literature \citep[e.g.,~][]{gizisetal2002, schmittliefke2004, silvestrietal2005, moutouetal2017}. If several measurements were taken for a given star, we have chosen the one with the strongest polarimetric signal. We analyze here only one phase at which the signal is maximal, that is, a snapshot of the star's rotational phase. The exposure times for the observed spectra are short compared to the rotational periods of the stars with values of about 0.7\%-6.0\% of the rotational periods (except for VB 8 where the exposure time is $\sim$38\% of the rotational period).  A next step will be an analysis of the rotational dependence of parameters. The basic information on the selected data is provided in Table~\ref{tab:tabdata}. These data were previously analyzed and presented in several papers (see references provided in the table caption).  

\section{Molecular and atomic data}\label{moldat}

The considered molecular species were previously described in \citet{aframberdyugina2015}, but for the sake of completeness, we briefly present them in this section again, with a more detailed overview of FeH, as it is a very sensitive diagnostic for M dwarfs. We also include magnetically sensitive  atomic lines, to determine surface temperatures and to measure magnetic fields.

\subsection{FeH at 8700 and 9900-10000\,\AA}
The molecule FeH is one of the most sensitive indicators of magnetic fields in cool stellar atmospheres. Here, we consider two different FeH bands of the Wing-Ford system F$^{4}\Delta$--X$^{4}\Delta$: the (1,0) band near 8700\,\AA\ and (0,0) band around 1$\mu$m. The magnetic sensitivity of the 1$\mu$m band  was clearly demonstrated in the sunspot atlas of \citet{wallaceetal1998}. A great advantage of using the Wing-Ford bands of FeH as a Zeeman diagnostic is the fact that numerous lines of the same species are visible in a relatively narrow wavelength region throughout spectra of M and L dwarfs, but they are less blended in contrast to other molecular features. In addition, many lines in these bands become noticeably broadened due to Zeeman splitting already at field strengths of 2$-$3\,kG.  Zeeman broadened FeH lines in an active M dwarf were detected by \citet{valentietal2001}. They employed the sunspot measurements by \citet{wallaceetal1998} to model the stellar spectrum, since a theoretical description of the FeH molecule was not available at that time. \citet{berdyuginaetal2001, berdyuginaetal2003} modeled synthetic Stokes profiles of FeH lines and showed the usefulness of the FeH F$^{4}\Delta$--X$^{4}\Delta$ system around 1$\mu$m for diagnosing solar and stellar magnetic fields, once the spin-coupling constants are available. 

The FeH F$^{4}\Delta$--X$^{4}\Delta$ system is produced by transitions between two electronic quartet states with $\Omega=\frac{7}{2},\frac{5}{2},\frac{3}{2}$, and $\frac{1}{2}$, $\Omega$ being the quantum number for the component of the total electronic angular momentum along the internuclear axis of a diatomic molecule. The coupling of the angular momenta is intermediate between limiting Hund's cases (a) and (b) \citep[see ][]{herzberg1950}. The necessary molecular constants, which describe how strongly the different internal angular momenta are coupled to each other, were unknown until provided by \citet{dulicketal2003}, allowing further analyses of this particular system.  Starting from the available Hamiltonian, \citet{aframetal2007,aframetal2008} calculated a perturbed molecular Zeeman effect, computed Land\'e factors of the energy levels and transitions, and calibrated molecular constants to account for an unknown perturbation by comparing observed and modeled polarized sunspot spectra.  Polarized spectra are affected more strongly by internal perturbations than intensity spectra. Hence, differences in the Land\'e factors can be better seen in polarization which makes this method more sensitive. In addition, polarization is important for determining the sign of the effective Land\'e factor, which can easily be negative in molecular transitions, as shown by \citet{berdyuginasolanki2002}. The progress in the theoretical description of this FeH system finally allowed \citet{aframetal2009} to start applying this theory for studying magnetic fields on M dwarfs. In this work, we fully exploit its diagnostic capabilities.  

The magnetic sensitivity varies strongly among the FeH lines of different rotational branches and total angular momentum $J$-numbers within the observed wavelength range. Transitions within P-, Q-, and R-branches have $\Delta J= -1, 0$, and $+1$, respectively. Our primary diagnostic was the highly sensitive FeH $Q_{3}$(2.5) line at 9975.476 \AA\ with the effective Land\'e factor $g_{\rm eff}=$0.96 \citep{aframetal2008}. Its Stokes I profile changes quickly as the magnetic field strength increases from 1 to 5 kG at  $T_{\mathrm{eff},2}=$3500K \citep[see ][]{aframetal2009}. The synthetic profile develops a characteristic, almost rectangular shape until it completely splits into Zeeman components. This particular line is therefore excellent for determining magnetic fields from intensity spectra. The $R_{1}$(2.5) and $R_{1}$(6.5) lines at 9976.397 \AA\ and 9978.721 \AA\ (with $g_{\rm eff}$ of 0.18 and 0.19, respectively) react only moderately to incremental changes of 1\,kG in the magnetic field strength and are almost insensitive to small variations of about 0.1\,kG \citep{aframetal2009}. A joint analysis of such lines with high and low magnetic sensitivity helps to disentangle the magnetic field strength and its filling factor. However, we emphasize that including too many marginally magnetically sensitive lines into an analysis may significantly dilute the result and increase uncertainties.

Previously, \citet{reinersbasri2006} investigated Zeeman broadening in FeH lines of early M dwarfs. They developed a technique to estimate the mean magnetic field from a comparison to reference spectra of early M dwarfs with magnetic field measurements calculated in atomic lines. \citet{reinersbasri2007} employed this method for  a sample of 22 M stars providing mean magnetic fields (averaged over the visible surface) for M dwarfs of spectral types M2--M9 with the accuracy of $\pm$1~kG. \citet{wendeetal2011} identified magnetically sensitive FeH lines in high resolution near-infrared CRIRES spectra through comparison of an active and inactive M dwarf. Laboratory measurements of the Zeeman response of Wing-Ford FeH lines were carried out by \citet{crozetetal2012, crozetetal2014}. Then, they used the newly determined  Land\'e factors to deduce the magnetic field in sunspots from Stokes V profiles. \citet{shulyaketal2010, shulyaketal2011a, shulyaketal2011b, shulyaketal2014}  used FeH to carry out  modeling and analysis of magnetic fields in selected M dwarfs with a method similar to \citet{reinersbasri2007}. 

\subsection{TiO at 7054\,\AA} 
Spectra of sunspots and cool stars show many TiO lines. The first astronomical detection of TiO was recorded in spectra of M dwarfs \citep{fowler1904} and first observations of TiO molecular bands from starspots was reported by \citet{vogt1979} for a K2 star. Using TiO bands for analyzing starspot properties was suggested by \citet{ramseynations1980}, which  allowed the spot area and temperature to be measured \citep[e.g.,~][]{neffetal1995, onealetal1996}. \citet{valentietal1998} tested the spectral synthesis of TiO lines for reproducing the optical spectrum of an inactive M dwarf. The first spectropolarimetric models and observations in sunspots were obtained by \citet{berdyuginaetal2000}. \citet{berdyuginaetal2003} concluded that the TiO $\gamma(0,0)R_3$ system band head at 7054\,\AA\ is one of the best molecular diagnostics of the magnetic field in sunspot umbra and starspots. First circular polarization signals in TiO lines were observed and modeled in three M dwarfs  \citep{berdyuginaetal2006b}. 

The triplet states of this system are in an intermediate Hund's case (a-b) and require accounting for perturbations for precise measurements \citep{berdyuginasolanki2002}. This is the approach we follow in this paper. Inferences of starspot temperature and filling factor on five highly active stars were shown by \citet{onealetal2004}, by fitting TiO bands using spectra of inactive G and K stars to represent the unspotted photospheres of the active stars and spectra of M stars to represent the spots.

\subsection{MgH at 5200\,\AA} 
The molecular Zeeman splitting for MgH was theoretically predicted by \citet{kronig1928} and \citet{hill1929}, while laboratory measurements of the Zeeman effect were analyzed in \citet{crawford1934}. \citet{laborde1961} described MgH spectral lines and  compared laboratory wavelengths with solar ones. MgH lines were employed for determining the surface gravity ($\log g$) of Arcturus \citet{belletal1985} and other cool stars \citep{berdyuginasavanov1992}. \citet{berdyuginaetal2000} reported the first spectropolarimetric measurements and model of MgH lines in sunspots. 

The A$^{2}\Pi$ state of the MgH A$^{2}\Pi$--X$^{2}\Sigma^{+}$ (0,0) transition is an intermediate Hund case (a-b) with spin-orbit constants of $Y =5.7$, i.e., approaching a pure Hund case (b). Th X$^{2}\Sigma^{+}$ state is pure case (b). Both states require proper perturbation calculations of the Paschen-Back effect. With this, the MgH A$^{2}\Pi$--X$^{2}\Sigma^{+}$ system at 5200\,\AA\ is a sensitive tool  for studying stellar magnetic \citep{berdyuginaetal2000, berdyuginaetal2005}.

\subsection{CaH at 6938\,\AA} 
Calcium monohydride (CaH) is an important astrophysical molecule, and its bands have been used as indicators of cool stars luminosities \citep{oehman1934, mould1976, mouldwallis1977, barbuyetal1993}. Because CaH absorption is an important opacity source in brown dwarfs \citep[e.g., ][]{burrowsetal2001, kirkpatrick2005}, CaH bands are valuable for studying brown dwarfs too. The CaH A$^{2}\Pi$--X$^{2}\Sigma^{+}$ band system is observed in the wavelength region at 6600--7600\,\AA. 

The ground state X$^{2}\Sigma^{+}$ is described by a pure Hund case (b), while the excited state A$^{2}\Pi$ is intermediate between Hund cases (a) and (b). In addition, the Paschen-Back effect must be considered for CaH \citep{berdyuginaetal2003}. \citet{berdyuginaetal2006a} obtained the first polarimetric observations and synthesis of CaH Stokes profiles and confirmed the diagnostic value of CaH  in cool astrophysical sources.

\subsection{Atomic lines} 
A few Zeeman sensitive atomic lines are also included in this study: two Fe~I lines at 8468.4\,\AA\ (effective  Land\'e factor $g_{\rm}$=2.50) and 8514.1\,\AA\ ($g_{\rm}$=1.83)  and three Ti~I lines at  8364.2\,\AA\ ($g_{\rm}$=1.43), 8377.9\,\AA\ ($g_{\rm}$=0.88), 8382.5\,\AA\ ($g_{\rm}$=1.25). The two last Ti~I lines are referred throughout the paper as 8380\,\AA\ lines. These lines were chosen because they cover a range of Land\'e factors, that helps disentangling the magnetic field strength from the area filled by that field (filling factor). Also, a number of weak blending molecular lines were included into the line list for computing synthetic spectra.

\section{Method}\label{method}

\begin{figure}
\centering
\setlength{\unitlength}{1mm}
\begin{picture}(130,80)
\put(0,27){\begin{picture}(0,0) \includegraphics{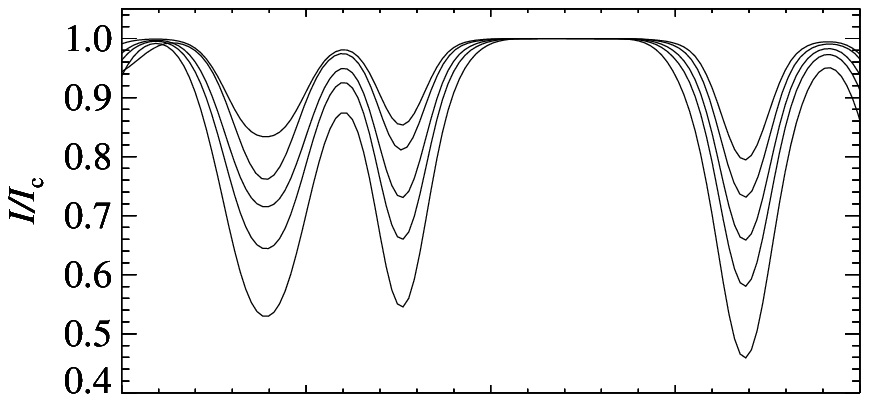} \end{picture}}
\put(0,-10){\begin{picture}(0,0) \includegraphics{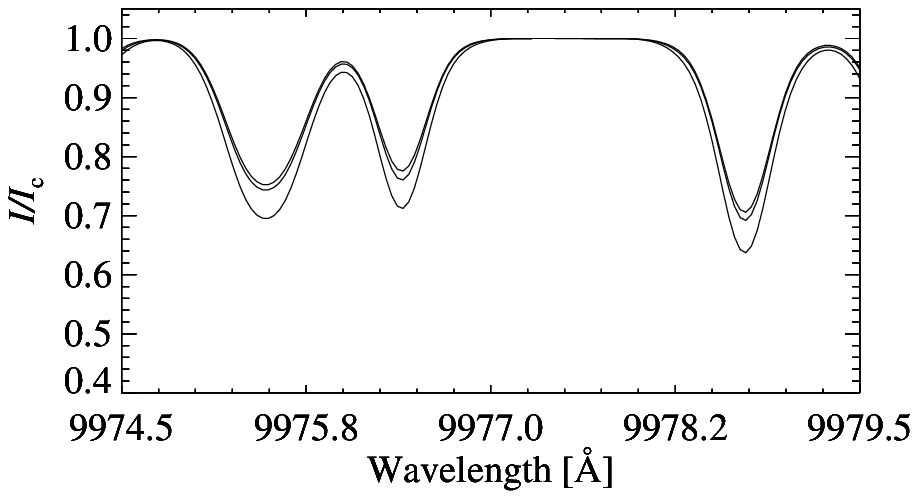} \end{picture}}
\end{picture}
\caption{Upper plot: Response of the highly magnetically sensitive FeH $Q_{3}$(2.5) line at 9975.476 \AA\  and the marginally sensitive FeH $R_{1}$(2.5) and $R_{1}$(6.5) lines at 9976.397 \AA\ and 9978.721 \AA, respectively, to changes of 200~K in the effective temperatures for the range of 2600~K to 3400~K. The line profiles become deeper with cooler temperatures. A magnetic field strength of 3000 G was applied and instrumental (0.3\,\AA) broadening. Lower plot: The same for changes of log g from 4.0 to 5.0 in steps of 0.5 and an effective temperature of 3000~K.}

\label{fig:vartemp}
\end{figure}

 Here, we present our method with which we determined effective temperatures of the photosphere and of magnetic features, magnetic field strengths and filling factors for nine M dwarfs (M1--M7). Our $\chi^2$ analysis is based on a comparison of observed and synthetic intensity and circular polarization profiles (Stokes I and V) of many magnetically sensitive  atomic and molecular lines. 

\subsection{Modeling Stokes profiles}\label{model}

We calculated synthetic molecular and atomic intensity and polarization profiles in the presence of magnetic fields using the code STOPRO \citep{solanki1987,solankietal1992,frutigeretal2000,berdyuginaetal2003,berdyuginaetal2005}. This code solves the polarized radiative transfer equations for all four Stokes parameters simultaneously under the assumption of the local thermodynamic equilibrium (LTE). The equations are solved for model atmospheres from \citet{allardetal2001} assuming solar element abundances. Both perturbed Zeeman and Paschen-Back effects were accounted for molecular transitions. Here, we use 1D models, which in the future can be replaced by 3D MHD models of starspots which are not yet available; even for sunspots, such models are still not fully realistic.

We took into account the following line-broadening effects: intrinsic, thermal (according to the local temperature in the model atmosphere), magnetic, microturbulent (1\,km/s), rotational, and instrumental (0.15\,\AA). For stars with the projected rotational velocity $v\sin i\ge5$\,km/s we considered $v\sin i$ values given by \citet{reinersbasri2007}. For stars with $v\sin i<5$\,km/s we adjusted $v\sin i$ values to obtain best fits. Uncertainties in $v\sin i<5$ result in negligible errors. For example, for the magnetic field strength, the error is smaller than 0.1\,kG.

For our analysis, we selected several spectral regions where various atomic and molecular transitions were observed. We divided them into three sets according to their temperature sensitivities:
\\
\textbf{"Atoms"} depicts all considered atomic lines, i.e., Fe I at 8468\,\AA\ and 8514\,\AA, and Ti I at 8364\,\AA\ and 8380\,\AA. 
\\
\textbf{"Mol1"} depicts the molecular lines of FeH at 9900\,\AA\ and 9975\,\AA\ and TiO at 7055\,\AA. In some cases, this set was further split into two subsets: one with the FeH lines only, the other only with the TiO lines, because TiO and FeH lines form at different heights in early M dwarf atmospheres.
\\
\textbf{"Mol2"} depicts the molecular lines of CaH at 6938\,\AA, MgH at 5200\,\AA, and FeH at 8700\,\AA.

The observed intensity (Stokes I) and circular polarization (Stokes V) profiles were used to determine parameters of M dwarf atmospheres. We employed a model with two components: a warm photosphere and cool spots, both being allowed to be magnetic. Cool magnetic regions are referred to as starspots, while warm magnetic regions are considered to be analogues to the solar photosphere network fields. This model can be described by five parameters as follows: 
\\
\indent $T_{\rm phot}$, photosphere temperature,\\
\indent $T_{\rm mag}$, magnetic region temperature,\\ 
\indent $B$, magnetic field strength,\\
\indent $f_{\rm I}$, magnetic field filling factor determined from Stokes I,\\ 
\indent $f_{\rm V}$, magnetic field filling factor determined from Stokes V.

Here, a filling factor $f$ is a fraction of the visible stellar surface area covered by regions with a magnetic field strength $B$ and temperature $T_{\rm mag}$. The fraction $1-f$ is the stellar surface with the photosphere temperature $T_{\rm phot}$ that is free from cool starspots or other magnetic regions. Because circular polarization of opposite signs for complex, spatially unresolved magnetic fields cancels out, we distinguish between filling factors determined from Stokes I and V. Accordingly, $B$ refers to a module (unsigned) of the magnetic field $|\vec B|$, when determined from only Stokes I, and to a residual (signed) line-of-sight (LOS) component $B_{\rm LOS}$, when determined from only Stokes V. When both Stokes I and V are fitted simultaneously (like in our model for most species), $B$ refers to the magnetic field module (unsigned) along the LOS. 

We characterize the magnetic field complexity (entanglement) as the difference between the filling factors determined from the intensity and polarization (here, Stokes V):
\begin{equation}
\epsilon_{\rm IV}=f_{\rm I} - f_{\rm V}. 
\end{equation}
This difference quantifies the fraction of the magnetic area on the star that is highly entangled and remains hidden if only Stokes V is analyzed.

With this model, we have searched for solutions with and without cool magnetic spots as well as with and without magnetic network fields. As described in the Sections~\ref{procedure}~and~\ref{red}, various combinations of these features have been found for different atomic and molecular species. For instance, atomic and mol2 lines have been found to be suitable to probe warmer magnetic regions (network) on M dwarfs, while mol1 lines are sensitive to cooler regions (starspots). 

We have also tested a three-component model, assuming the existence of a non-magnetic photosphere, magnetic areas in the photosphere (network) and magnetic cool spots (starspots) simultaneously. Results obtained with such a three-component model and the two-component model described above were very similar.

\begin{table*}
\caption{Summary of iterative $\chi^{2}$ minimization procedure used in this paper. The sets of spectral features (atoms, mol1, mol2) are described in Sect.~\ref{model}. The results are presented in Table~\ref{tab:tabmol}  and shown in Fig.~\ref{fig:tabmol}}. 
\smallskip
\label{tab:tabproc}
\centering
\begin{tabular}{lcllll}
\\
\hline
\noalign{\smallskip}
Species &Used   &Assumed &Retrieved &Range, step & Probed atmosphere features \\
        &Stokes &        &          &            & \\
\hline
Ti I 8380\,\AA\ & I &$B$=0         &$T_{\rm phot}$(Ti)& 2000K--4000K, 100K& An average photosphere \\
                &   &              &                  &                   & with the highest temperature.\\
\hline
mol1        & I &$T_{\rm phot}$(Ti)                  &$T_{\rm mag}$ &2000K--4000K, 100K& Starspots: TiO and FeH lines\\ 
            &   &$T_{\rm mag}$ $\leq$ $T_{\rm phot}$&$f_{\rm I}$    &0.0--1.0, 0.1     & were fitted separately  \\ 
            &   &                                    &               &                 & (form at different heights).\\
\hline
mol2        & I &$T_{\rm phot}$(Ti)                  &$T_{\rm mag}$ &2000K--4000K, 100K& Magnetic network. \\ 
            &   &$T_{\rm mag}$ $\leq$ $T_{\rm phot}$ &$f_{\rm I}$    &0.0--1.0, 0.1     &  \\ 
\hline
atoms       &I, V & $T_{\rm phot}$=$T_{\rm mag}$   & $B$         & 0G--7000G, 100G & Magnetic network. \\
            &     &                                & $f_{\rm V}$ & 0.0--0.2, 0.01  & \\
\hline
mol1        &I, V & $T_{\rm phot}$(Ti) &$B$         & 0G--7000G, 100G  & Starspots.\\
            &     & $T_{\rm mag}$      &$f_{\rm V}$ & 0.0--0.2, 0.01   & \\
			&     & $f_{\rm I}$        &            &                  & \\			
\hline
mol2        & I   & $T_{\rm phot}$(Ti) &$B$         & 0G--7000G, 100G  & Magnetic network. \\
            &     & $T_{\rm mag}$      &            &                  & The Stokes V signals \\
			&     & $f_{\rm I}$        &            &                  & are too weak and not used.\\
\hline
\end{tabular}
\end{table*}

\begin{figure*}
\centering
\setlength{\unitlength}{1mm}
\begin{picture}(145,160)   
\put(-17,100){\begin{picture}(0,0) \includegraphics{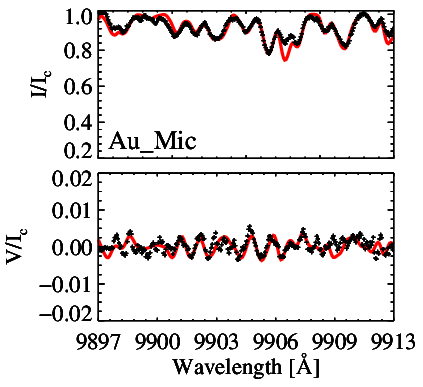} \end{picture}}
\put(43,100){\begin{picture}(0,0) \includegraphics{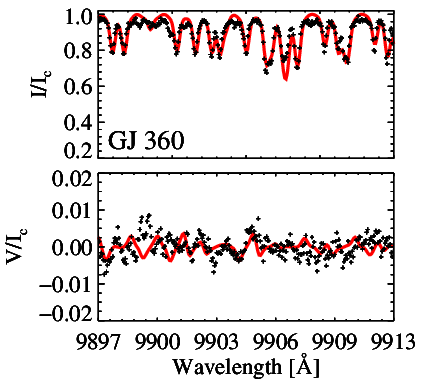} \end{picture}}
\put(103,100){\begin{picture}(0,0) \includegraphics{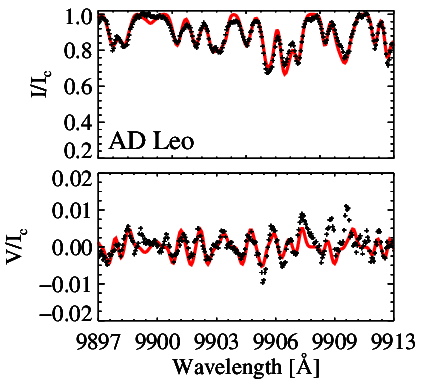} \end{picture}}
\put(-17,46){\begin{picture}(0,0) \includegraphics{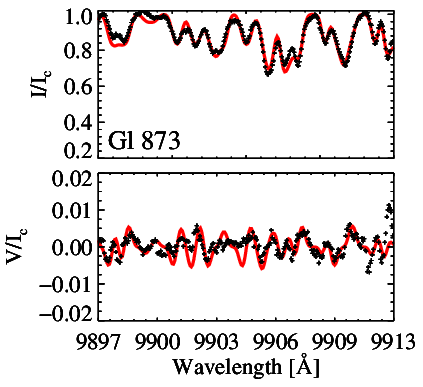} \end{picture}}
\put(43,46){\begin{picture}(0,0) \includegraphics{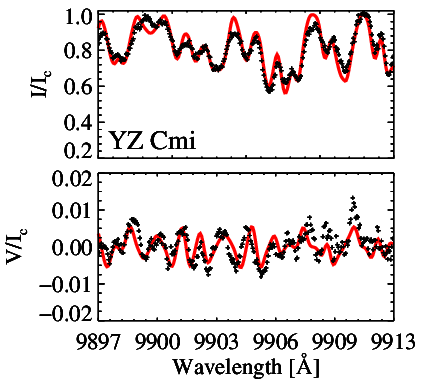} \end{picture}}
\put(103,46){\begin{picture}(0,0)  \includegraphics{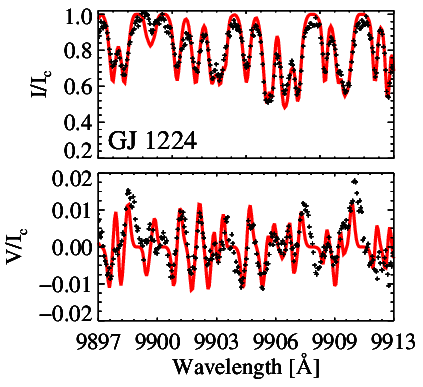} \end{picture}} 
\put(-17,-8){\begin{picture}(0,0) \includegraphics{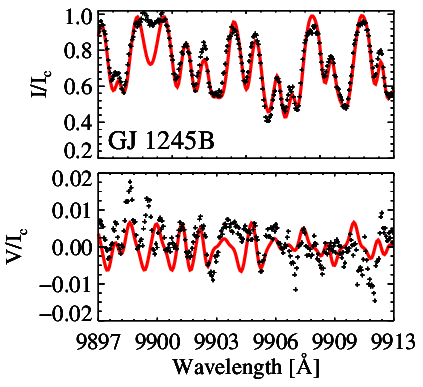} \end{picture}} 
\put(43,-8){\begin{picture}(0,0)  \includegraphics{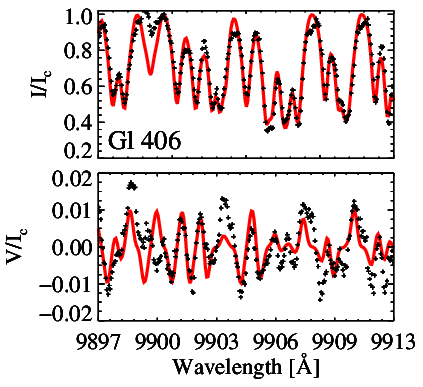} \end{picture}}
\put(103,-8){\begin{picture}(0,0) \includegraphics{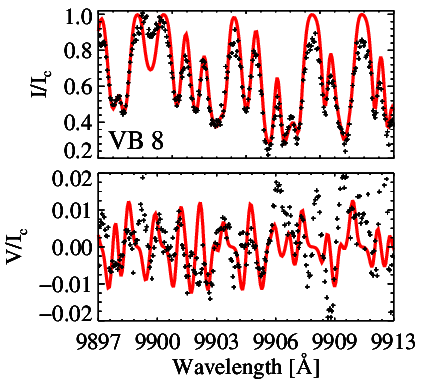} \end{picture}}
\end{picture}
\caption{This illustrates the high temperature and magnetic sensitivity of FeH lines which is reproduced by our models. Observed (black) and synthetic (red) intensity and circular polarization profiles for FeH at 9897-9913 \AA\ in the considered sample of M dwarfs. The synthetic profiles were calculated using two components: one non-magnetic (''photospheric'') and one magnetic (''spot'')  atmosphere model \citep{allardetal2001} with best-fit parameters obtained from fits of Stokes I and V spectra for FeH in the considered wavelength region only. All the following parameters were obtained simultaneously in one $\chi^{2}$ minimization: the photosphere and spot temperatures, the magnetic field strength, the spot filling factor for the intensity $f_{\rm I}$, and the circular polarization $f_{\rm V}$.}
\label{fig:bestfitfeh}
\end{figure*}

\begin{figure*}
\centering
\setlength{\unitlength}{1mm}
\begin{picture}(145,160)   
\put(-17,100){\begin{picture}(0,0) \includegraphics{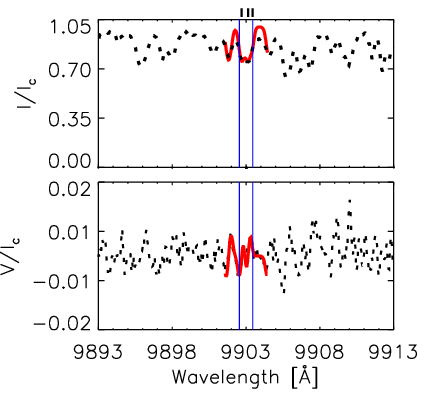} \end{picture}}
\put(28,100){\begin{picture}(0,0) \includegraphics{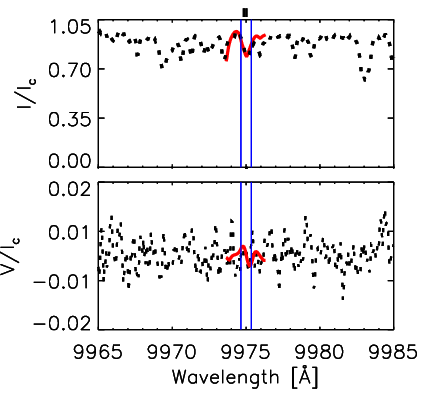} \end{picture}}
\put(73,100){\begin{picture}(0,0) \includegraphics{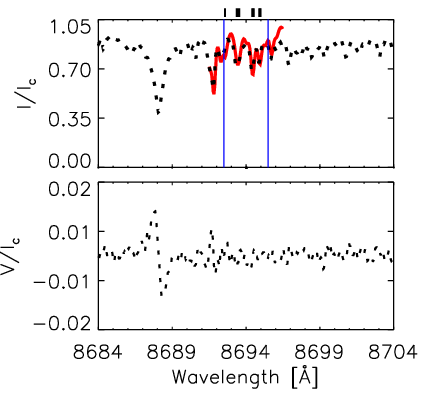} \end{picture}}
\put(118,100){\begin{picture}(0,0)  \includegraphics{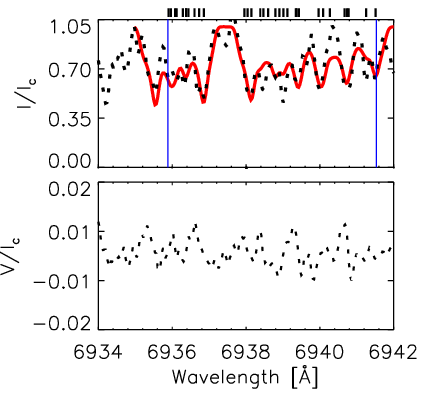} \end{picture}} 
\put(-17,47){\begin{picture}(0,0) \includegraphics{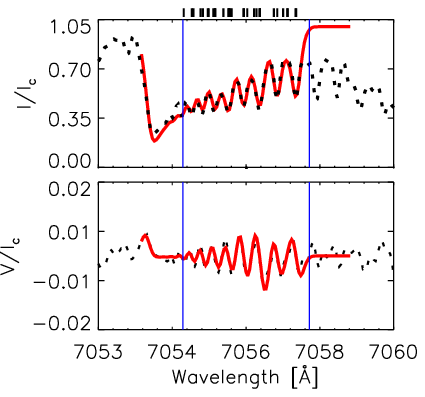} \end{picture}} 
\put(28,47){\begin{picture}(0,0)  \includegraphics{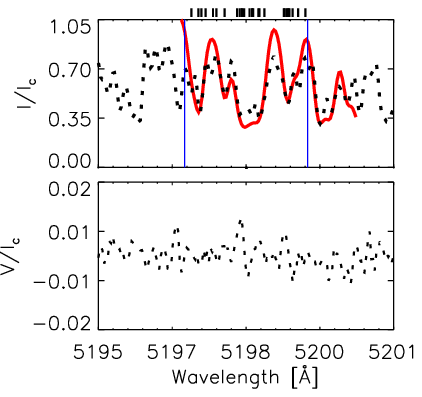} \end{picture}}
\put(73,47){\begin{picture}(0,0) \includegraphics{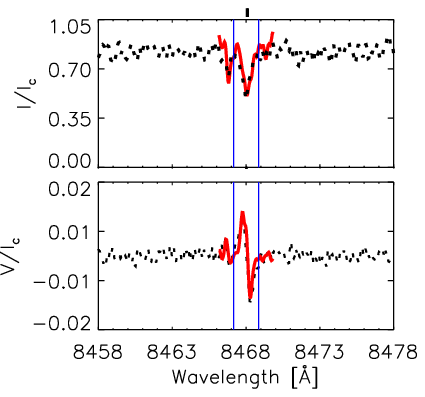} \end{picture}}
\put(118,47){\begin{picture}(0,0) \includegraphics{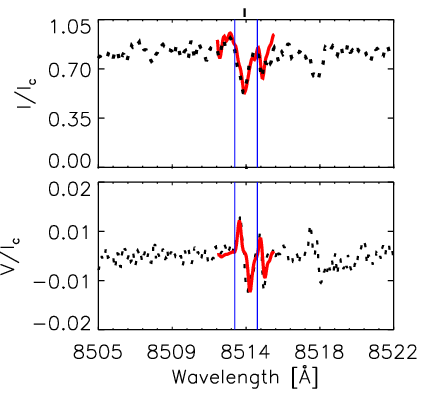} \end{picture}}
\put(28,-6){\begin{picture}(0,0) \includegraphics{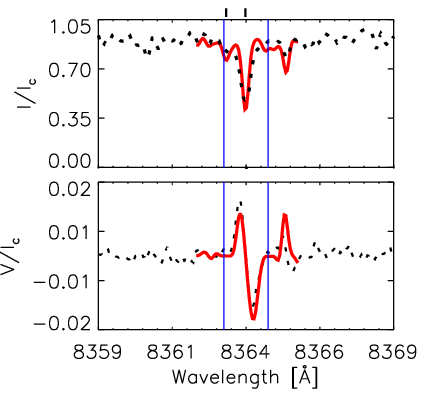} \end{picture}}
\put(73,-6){\begin{picture}(0,0) \includegraphics{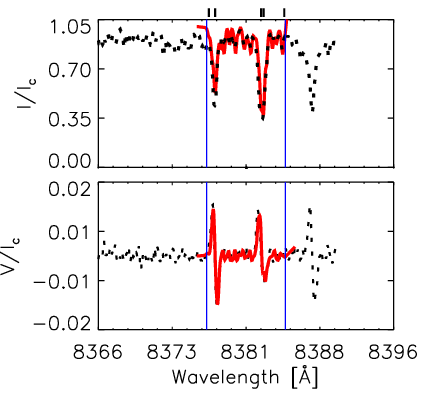} \end{picture}}
\end{picture}
\caption{Observed (black) and synthetic (red) intensity and circular polarization profiles for AD Leo in all considered wavelength regions. The synthetic profiles were calculated using two components: one non-magnetic (''photospheric'') and one magnetic (''spot'')  atmosphere model \citep{allardetal2001}. Fits to Stokes I and V spectra for each atom and molecule were obtained individually (see Sect.~\ref{procedure}), but all the parameters were obtained simultaneously: the photosphere and spot temperatures, the magnetic field strength, the spot filling factor for the intensity $f_{\rm I}$, and the circular polarization $f_{\rm V}$. In three regions Stokes V was not fitted because of lower signal to noise in this particular wavelength region and observation, while in other regions the signal is stronger, despite the fact that it looks like noise because the region contains many lines.}
\label{fig:bestfitadleo}
\end{figure*}

\begin{table*} 
\caption{Best fit parameters (including sensitivities as determined in  Sect.~\ref{error}) for all considered stars obtained from the $\chi^{2}$ analyses. For molecules (sets mol1 and mol2), $T_{\rm phot}$ is fixed to $T_{\rm phot}$(Ti).  For atoms (set atoms), $T_{\rm phot}=T_{\rm mag}$.}

\smallskip
\label{tab:tabmol}
\centering
\begin{tabular}{llllllll}
\\
\hline
\noalign{\smallskip}
Star    &Species  &$T_{\rm phot}$[K]&$T_{\rm mag}$[K]&$B$[kG]&$f_{\rm I}$&$f_{\rm V}$& $\epsilon_{\rm IV}$\\
sensitivity&      &$\pm$50~K&$\pm$50~K&$\pm$0.5~kG&$\pm$0.1&$\pm$0.02 & \\
\hline
AU Mic  &  Ti I   &  3700  &  -   &  -    &  -    & -    & -   \\
        &  atoms  &  3800  &  -   &  4.3  &  0.8  & 0.01 & 0.79\\
        &  mol1   &  -     &  3700&  3.7  &  0.1  & 0.01 & 0.09\\
        &  mol2   &  -     &  3700&  3.2  &  1.0  & -    & -   \\
GJ 360  &  Ti I   &  3800  &  -   &  -    &  -    & -    & -   \\		
        &  atoms  &  3800  &  -   &  5.1  &  0.2  & 0.01 & 0.19\\
        &  mol1   &  -     &  2000&  4.5  &  0.7  & 0.01 & 0.69\\ 
        &  mol2   &  -     &  3800&  5.0  &  1.0  & -    & -   \\
AD Leo  &  Ti I   &  3700  &  -   &  -    &  -    & -    & -   \\
        &  atoms  &  3700  &  -   &  3.5  &  0.9  & 0.05 & 0.85\\
        &  mol1   &  -     &  2200&  5.6  &  0.7  & 0.05 & 0.65\\ 
        &  mol2   &  -     &  3700&  4.3  &  0.4  & -    & -   \\
EV Lac  &  Ti I   &  3600  &  -   &  -    &  -    & -    & -   \\		
        &  atoms  &  3600  &  -   &  5.1  &  0.7  & 0.03 & 0.67\\
        &  mol1   &  -     &  2000&  5.6  &  0.7  & 0.04 & 0.66\\
        &  mol2   &  -     &  3600&  6.8  &  0.8  & -    & -   \\
YZ Cmi  &  Ti I   &  3400  &  -   &  -    &  -    & -    & -   \\
        &  atoms  &  3400  &  -   &  4.3  &  1.0  & 0.01 & 0.99\\
        &  mol1   &  -     &  2300&  4.4  &  0.6  & 0.01 & 0.59\\
        &  mol2   &  -     &  3400&  5.0  &  1.0  & -    & -   \\
GJ 1224 &  Ti I   &  3500  &  -   &  -    &  -    & -    & -   \\		
        &  atoms  &  3500  &  -   &  4.2  &  0.8  & 0.08 & 0.72\\
        &  mol1   &  -     &  2100&  4.5  &  0.8  & 0.20 & 0.60\\
        &  mol2   &  -     &  2000&  5.0  &  0.7  & -    & -   \\
GJ 1245B&  Ti I   &  3200  &  -   &  -    &  -    & -    & -   \\		
        &  atoms  &  3400  &  -   &  4.3  &  1.0  & 0.02 & 0.98\\
        &  mol1   &  -     &  2300&  5.7  &  0.7  & 0.02 & 0.68\\
        &  mol2   &  -     &  3200&  5.0  &  1.0  & -    & -   \\
CN Leo  &  Ti I   &  3100  &  -   &  -    &  -    & -    & -   \\		
        &  atoms  &  3400  &  -   &  4.5  &  1.0  & 0.08 & 0.92\\
        &  mol1   &  -     &  2000&  5.0  &  0.7  & 0.05 & 0.65\\
        &  mol2   &  -     &  3100&  5.0  &  1.0  & -    & -   \\
VB 8    &  Ti I   &  3100  &  -   &  -    &  -    & -    & -   \\			
        &  atoms  &  4000  &  -   &  7.0  &  1.0  & 0.20 & 0.80\\
        &  mol1   &  -     &  3100&  1.0  &  0.6  & 0.01 & 0.59\\
        &  mol2   &  -     &  3100&  5.0  &  1.0  & -    & -   \\
\end{tabular}
\end{table*}

\begin{figure*}
\centering
\resizebox{\hsize}{!}{\includegraphics{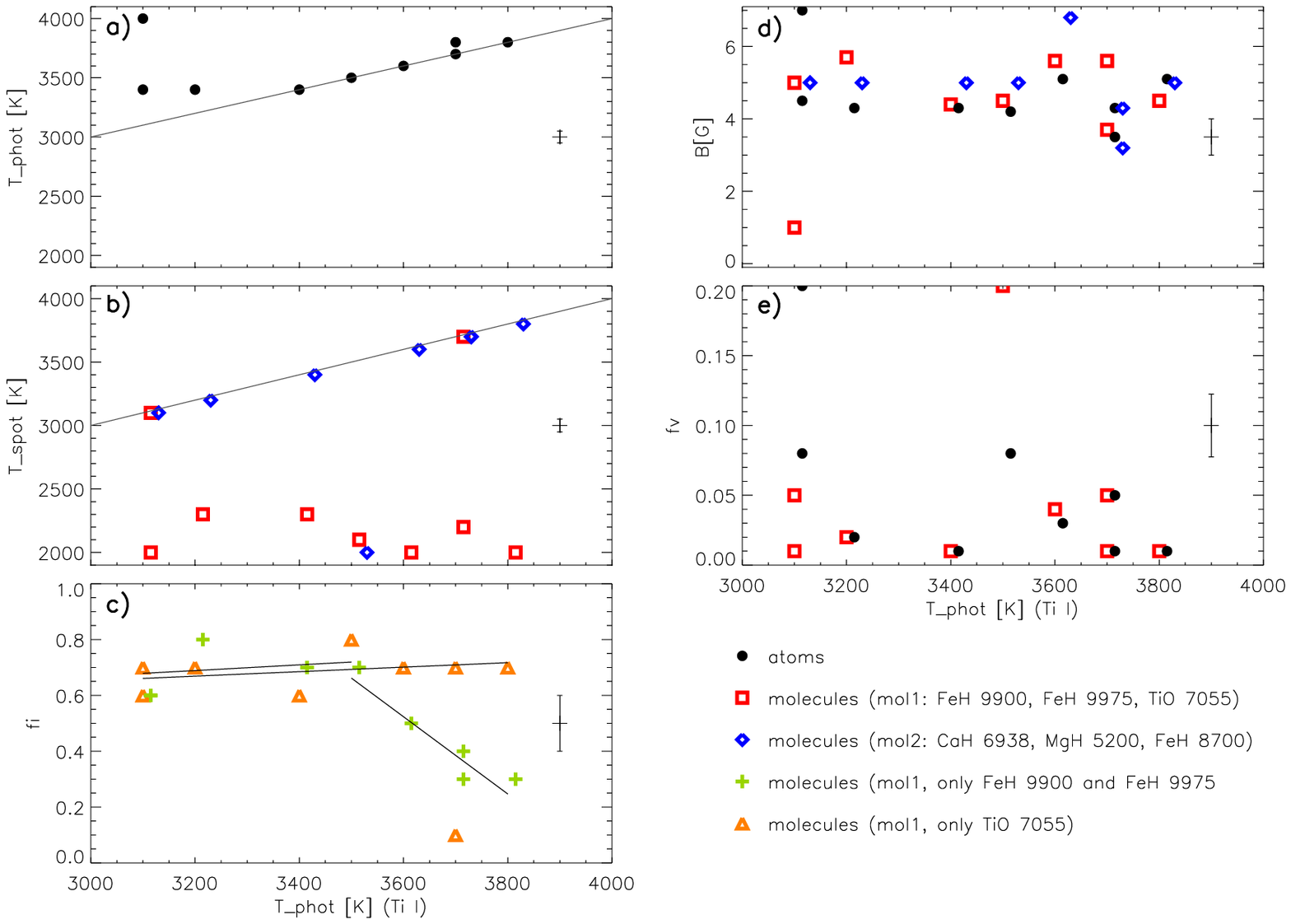}}
\caption{Temperature and magnetic field best-fit parameters (see Table~\ref{tab:tabmol}) for different magnetic regions on M dwarfs which are probed by five groups of spectral species.  $T_{\rm mag}$ and $f_{\rm I}$ are obtained from fitting Stokes I profiles only; $B$ is from fits of Stokes I and V together; $f_{\rm V}$ is from fits to Stokes V only; for the atoms, $T_{\rm phot}=T_{\rm mag}$ (see text for details). To avoid overlaps, some symbols are slightly shifted in the horizontal direction. Uncertainties of the parameters are indicated as thin-line crosses in each panel. The linear fits in panel \ref{fig:tabmol}c are described in Sect.~\ref{sec:ff}.} 
\label{fig:tabmol}
\end{figure*}

Since metal hydrides are sensitive to both stellar temperature and luminosity (see Sect.~\ref{moldat}), we used FeH lines for evaluating how the model atmosphere effective temperature $T_{\rm eff}$ and surface gravity $\log g$ affect the line profiles (Fig.~\ref{fig:vartemp}). We found a stronger response to $T_{\rm eff}$ changes for cooler stars. For $\log g$, the difference between 4.0 and 4.5 is negligible, but it is somewhat larger between 4.5 and 5.0, i.e., also for lower-mass, cooler stars. Considering uncertainties in spectral classification of our targets, we fixed $\log g = 4.5$ and then determined effective temperatures. In this case, a $\log g$ uncertainty of 0.1 dex would lead to errors in $T_{\rm eff}$ of $\sim$30\,K at an effective temperature of 3000\,K (see also discussion of our results in Sect.~\ref{red}).

\subsection{Parameter fitting procedure}\label{procedure}

We determined the model parameters introduced in the previous section for the observed stars using a $\chi^{2}$ minimization analysis by fitting the spectral range for each considered set.  

Since the model parameters are convolved, we used an iterative procedure. First, we attempted to formally fit all selected sets of atomic and molecular species simultaneously to obtain all the parameters at once. However, we found that the selected parameter set is insufficient for describing all the data, since the parameters refer to a single layer in the atmosphere and do not account for its complexity in height. In contrast, our data carry this height dependent information and, thus, cannot be fitted with parameters describing only one layer within the atmosphere. In addition, magnetic field parameters determined from polarization are influenced by the field entanglement via cancellations. In principle, if the model is complex enough, fitting all parameters at once should be possible, especially when including profile variability due to stellar rotation, i.e., similar to ZDI but in 3D. This is however beyond the scope of the present paper.

Therefore, we adapted our procedure.  We have fitted various combinations of the model parameters at the same time using different sets of data, thus, attempting to probe both the entire atmosphere and its different layers separately. We arrived at the most optimal procedure for our analysis, a summary of which is given in Table~\ref{tab:tabproc}. 

We first determined $T_{\rm phot}$ assuming $B=0$ from the Ti I 8380\,\AA\ Stokes I two line profiles. These lines were found to indicate systematically higher temperatures than other species. Therefore, it was fixed as a reference temperature for a non-magnetic photosphere in the subsequent analysis. Since in reality it fits a mix of the photosphere and spot contributions, we actually estimate a mean effective temperature, which could be 100--200K lower than the true photospheric temperature. On the other hand, network magnetic regions could be hotter than the photosphere (like on the Sun) which may help counteract the temperature bias. After fixing $T_{\rm phot}$, we iteratively determined the remaining model parameters from Stokes I and V profiles of all species as presented in Table~\ref{tab:tabproc}. We have found that atomic Stokes I and V profiles can be well fitted with a model assuming no cool spots but allowing for network magnetic regions. These model results are listed in Table~\ref{tab:tabmol} under the entry "atoms".

Examples of best fits to the observed Stokes profiles are shown in Fig.~\ref{fig:bestfitfeh} and Fig.~\ref{fig:bestfitadleo}: the first shows fits for FeH lines only but in all stars, and the second shows fits to spectra for all atomic and molecular species but only for the star AD Leo. For illustration purpose only, the best-fit synthetic spectra were calculated for each atom or molecule individually (since the analysis of all species together does not necessarily yield the best fit for the individual species), and with all the five model parameters obtained at once.  The fit in Fig.~\ref{fig:bestfitfeh} was obtained including only FeH lines without other blends. Hence, it gives an impression of the dominance of the FeH F$^{4}\Delta$--X$^{4}\Delta$ system throughout the extracted wavelength range of about 10 \AA. Poorer fits in Stokes V result from lower signal to noise, low or very complex signals, and cancellations in the circular polarization.

\subsection{Analysis of sensitivities}\label{error}

The sensitivities of the derived parameters were estimated from the average signal-to-noise ratio (S/N) level of the spectral region as listed in the online data archive files. The sensitivities correspond to the internal error of each parameter, when all other parameters are fixed. The $1\sigma$ errors of all the parameters (photosphere and spot temperatures, filling factors in intensity and circular polarization and field strength) have been estimated from the unnormalized $\Delta\chi^{2}$ level for the respective number of degrees of freedom. Sensitivities are given in Table ~\ref{tab:tabmol}. They do not include systematic errors due to missing blends and imperfections of the fits. Those systematic errors are difficult to determine because of the limited number of the parameters considered for describing complex spectra of M dwarfs.

\section{Results}\label{red}

Main results following the procedure described in Sect.\ref{procedure} comprise temperatures of the photosphere and magnetic features, magnetic field strengths, and magnetic filling factors in intensity and circular polarization for all nine stars and the atomic (atoms) and molecular (mol1 and mol2) sets. They are presented in Table~\ref{tab:tabmol}, visualized in  Fig.\ref{fig:tabmol}, and discussed here.

\subsection{The photosphere temperature}\label{tphot}

As noted above, the photosphere temperature was evaluated from the intensity  profiles of the two Ti I 8380\,\AA\ lines as the highest inferred value among all species. We conclude that these lines are least contaminated by cool spots (which are approximately 2200 K cooler than the photosphere), while they can still be contaminated by hot areas (which are about 200 K hotter). The evaluated photosphere temperature was fixed for the analysis of other line sets. It changes from 3800\,K for an M1 dwarf to 3100\,K for an M7 dwarf. These are somewhat higher than average effective temperatures (photosphere + cool spots) of M dwarfs determined from observed colors or spectral fitting without accounting for cool spots (3700\,K for M1 to 2700\,K for an M7  \citep[e.g., ][]{rajpurohitetal2013}. This discrepancy is explained by our finding that spots on M dwarfs are significantly cooler than the photosphere and can cover a large surface area (see Sect.~\ref{tspot} and~\ref{sec:ff}).

Interestingly, when we fitted $T_{\rm phot}$ using all selected atomic lines, we found even higher values for the three coolest M dwarfs (Fig.~\ref{fig:tabmol}a and third column in Table~\ref{tab:tabmol}). Among other possible explanations, this may indicate that network regions are significantly hotter than the photosphere on the coolest dwarfs, while we assumed they are of the same temperature as the photosphere.  When we attempted to include such features into our analysis (three-component model), the results were redundant. After $T_{\rm phot}$ was determined as described here, its value was fixed to obtain the other four parameters of our model.

\subsection{The temperature of magnetic regions}\label{tspot}

Fitting Stokes I and V of the selected Fe I and Ti I lines using four free model parameters ($T_{\rm phot}$ was fixed to $T_{\rm phot (Ti)}$) resulted in $T_{\rm mag}=T_{\rm phot}$ and a strong magnetic field. Thus, these lines probe a warm, highly magnetized photosphere, i.e., network. In fact, we have not found any lines that could be formed in a non-magnetized atmosphere on these M dwarfs.

When fitting Stokes I spectra of the CaH, MgH, and FeH 8700\,\AA\ lines (set mol2; their Stokes V spectra were not included because of low S/N) using the three model parameters (excluding $f_{\rm V}$), we found that the resulting $T_{\rm mag}$ matched well $T_{\rm phot}$ obtained with the Ti I 8380\,\AA\ lines (Fig.\ref{fig:tabmol}b) for all but two stars. This indicates that the mol2 set lines also form preferentially in the warm photosphere on M dwarfs, in contrast to G and K dwarfs where these lines are excellent diagnostics of cool starspots \citep{aframberdyugina2015}. Hence, the mol2 set CaH, MgH and FeH 8700\,\AA\ lines can be used to determine $T_{\rm phot}$ of M dwarfs and, therefore, provide an independent justification for our choice of $T_{\rm phot}$. Also, as is the case for atomic lines, they reveal that M dwarf photospheres are highly magnetized (see Sect.~\ref{bmag}).

Fitting Stokes I and V of the mol1 set TiO and FeH lines using the four free model parameters revealed for most stars $T_{\rm mag}$ between 2000--2300\,K with an average of 2100\,K, which is approximately constant across the M1 to M7 spectral classes (Fig.\ref{fig:tabmol}b). These cool spots are also highly magnetic (see Sect.~\ref{bmag}). Here, the difference between $T_{\rm phot}$ and $T_{\rm mag}$ reduces on average from 1700\,K for an M1 dwarf to about 1000\,K for an M7 dwarf. Interestingly, this range of spot temperature differences is typical for sunspots of various sizes \citep[e.g.,][]{berdyugina2005}. For a young, active M8.5 brown dwarf, \citet{berdyuginaetal2017} and \citet{kuzmychovetal2017} determined $T_{\rm phot}=2800$\,K and $T_{\rm mag}=2200$\,K using also atomic and molecular lines. This result agrees with the found trend and extends our finding to ultra-cool dwarfs.

The spot temperature difference with respect to the photosphere determined here for M dwarfs differs from that found from broad-band photometric measurements, as summarized by \citet{berdyugina2005}. This is probably due to a height-dependence of spot parameters as well as the higher sensitivity of molecular lines to the temperature than that of photometry (see Sect.~\ref{conc}).

In summary, thanks to the variety of our spectral diagnostics, we found two kinds of magnetic regions with clearly different temperatures in M dwarf atmospheres as a function of the spectral class: one is cold (starspots), and the other is warm (network).

\subsection{The magnetic field strength}\label{bmag}

The magnetic field strength $B$ was determined during the same fitting procedure as described for $T_{\rm mag}$ in Sect.~\ref{tspot}. It corresponds therefore to an average field strength of a given magnetic structure at a given layer of the atmosphere, as defined by the corresponding species.  

The $B$ values found from atomic and molecular lines are confined to the range of 3--6\,kG (there are only three outliers), with an average of about 5\,kG (Fig.~\ref{fig:tabmol}d). This average is practically constant throughout the considered spectral class range of M1--M7 dwarfs. \citet{berdyuginaetal2017} and \citet{kuzmychovetal2017} similarly found $B\sim$5$\,$kG in their study of a young, active M8.5 brown dwarf. 

In summary, a cooler starspot atmosphere probed by the mol1 set and a hotter network atmosphere probed by the atoms and mol2 set are both magnetic, indicating highly magnetized atmospheres on these stars. The constant trend for the magnetic field implies that magnetic structures probably become locally smaller in scale toward lower mass M dwarfs.
 
\subsection{The magnetic field filling factors}\label{sec:ff}

\begin{figure}
\centering
\setlength{\unitlength}{1mm}
\begin{picture}(145,120)  
\put(0,60){\begin{picture}(0,0) \includegraphics{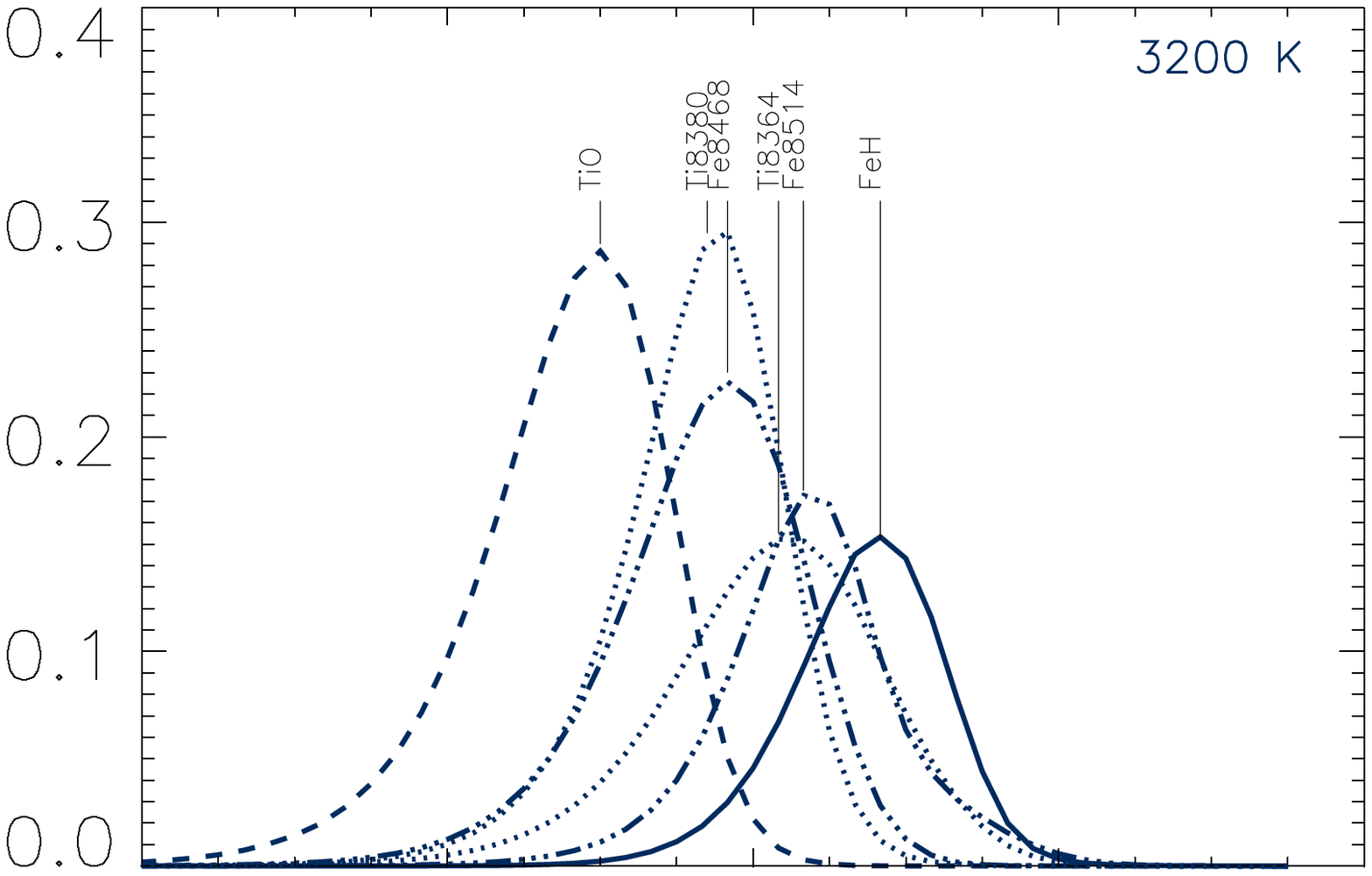} \end{picture}}
\put(0,7){\begin{picture}(0,0) \includegraphics{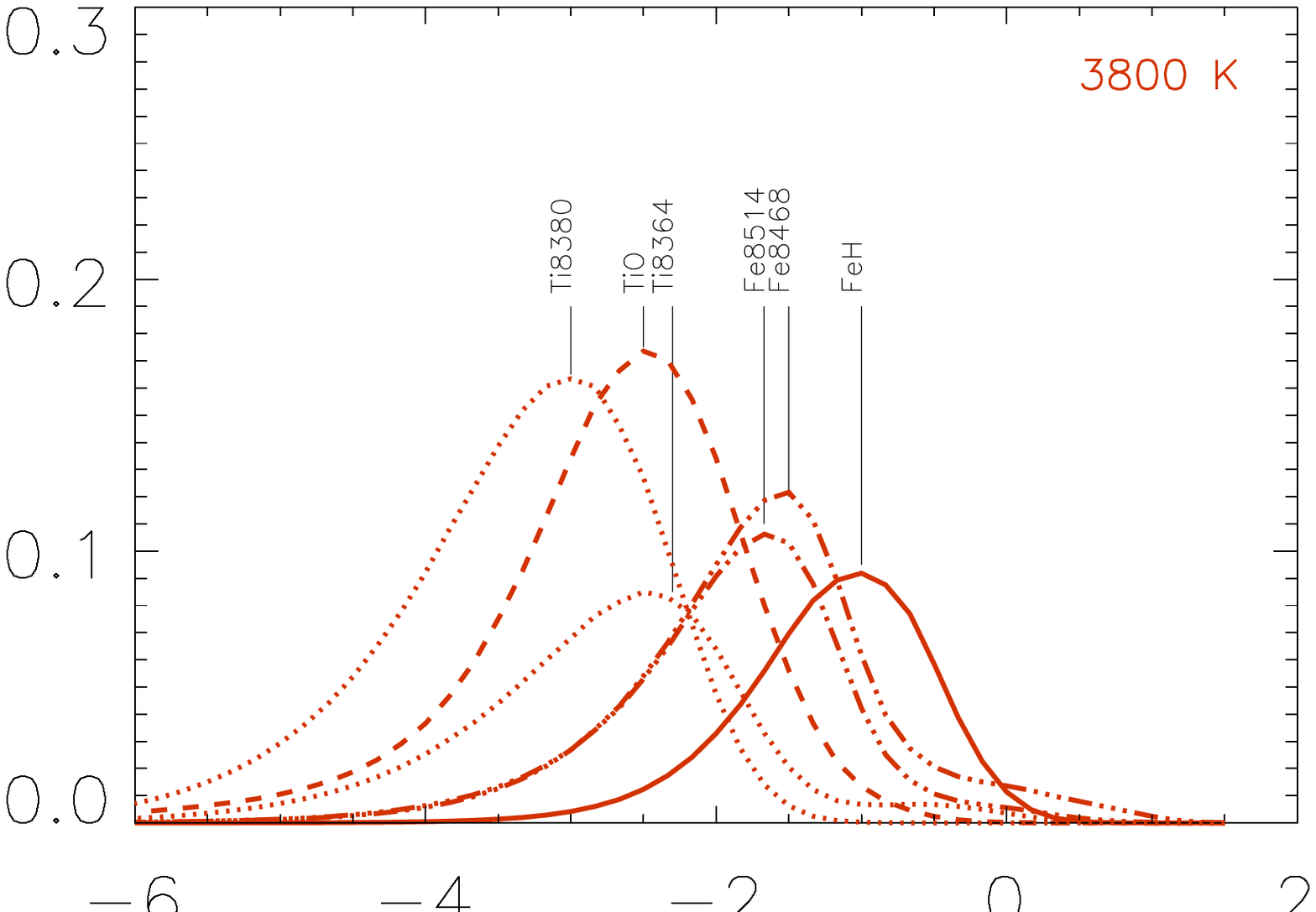} \end{picture}}
\end{picture}
\caption{Contribution functions for selected atomic and molecular species for two atmosphere models with effective temperatures 3200K and 3800K versus the optical depth at 1.2\,$\mu$m. Maxima indicate effective formation heights of the lines.}
\label{fig:cf}
\end{figure}

\begin{figure*}
\centering
\includegraphics[width=190mm]{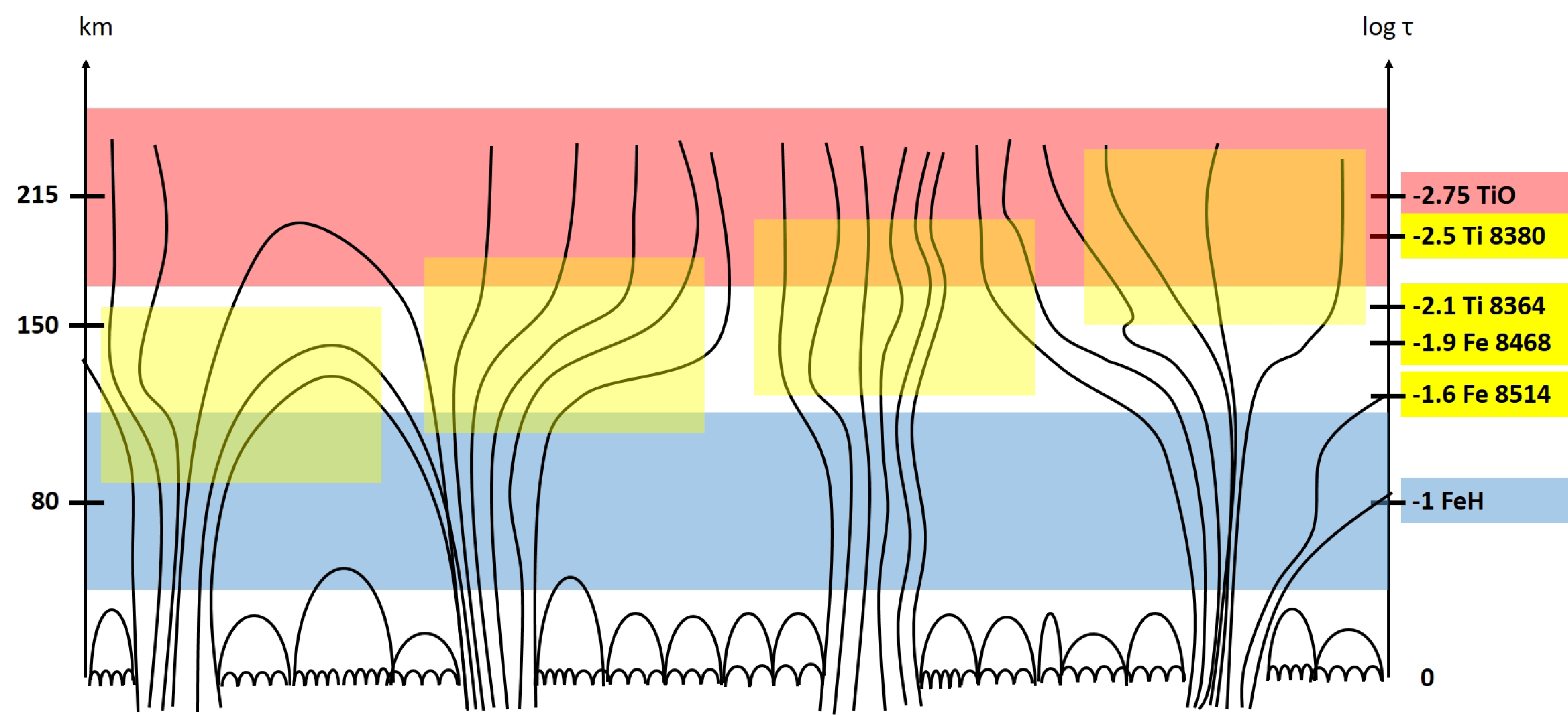}
\caption{A sketch of different magnetic regions in M dwarf atmospheres. Starspots expand toward higher layers and have a large filling factor in the upper atmosphere. Small-scale magnetic fields (network) dominate the lower atmospheres, also with a high filling factor. The geometrical scale given in axis on the left (km) belongs to the field lines. Atomic and molecular lines are formed at different optical depths (emphasized by different colors; the optical depth scale is given in the axis on the right) which allows us to probe changing magnetic strengths and structure size at different heights. The atmosphere parameters used for the sketch: effective temperature of 3500K and magnetic field strength of 5 kG. Inside the magnetic spots, the temperature is about 2200K, which is not shown here.}
\label{fig:skizze}
\end{figure*}

The magnetic field filling factors $f_{\rm I}$ and $f_{\rm V}$ were determined during the same fitting procedure as described for $T_{\rm mag}$ in Sect.~\ref{tspot}. The filling factor $f_{\rm V}$ has been determined for the atom and mol1 sets (Fig. \ref{fig:tabmol}e), while for the mol2 set, the circular polarization signals were too weak when compared to the noise.

For the atom and mol2 sets, the photosphere and magnetic region temperatures were found to be the same, except for one case (see Sect.~\ref{tspot}). Thus, to distinguish the non-magnetic photosphere and network, we analyzed atomic Stokes I and V simultaneously. This resulted in very high $f_{\rm I}$ (between 0.7 and 1.0, except for one star) and very low $f_{\rm V}$ (0.01 to 0.08, with only one star having 0.2). This large difference between $f_{\rm I}$ and $f_{\rm V}$ indicates a strong entanglement of the network magnetic field in M dwarfs. This also explains why the mol2 set lines have very low Stokes V signals.

For the mol1 set, we analyzed results for FeH and TiO separately. These molecules reveal cool magnetic starspots (see Sect.~\ref{tspot}) with $f_{\rm I}$ up to 0.8. Figure \ref{fig:tabmol}c shows that the FeH $f_{\rm I}$ increases from 0.3 to 0.8 for earlier to later M dwarfs, while the TiO $f_{\rm I}$ remains on average constant at about 0.7 (except for one star).  
We computed linear fits with an outlier-resistant two-variable linear regression for the FeH and TiO values separately for $T_{\rm eff}$ $\le$ 3500 K and for $T_{\rm eff}$ $\ge$ 3500 K. The parameters for TiO and for FeH (for $T_{\rm eff}$ $\le$ 3500 K) are: $f_{\rm I, TiO}$ = (0.00008$\pm$0.000088)$*T_{\rm eff}$+(0.41$\pm$0.30) and $f_{\rm I, FeH,}$ = (0.0001$\pm$0.00026)$*T_{\rm eff}$+(0.36$\pm$0.86). The parameters for FeH (for $T_{\rm eff}$ $\ge$ 3500 K) are: $f_{\rm I, FeH,}$ = (-0.001$\pm$0.0002)$*T_{\rm eff}$+(5.5$\pm$0.9). The errors of the linear fits show that the difference between the trends is significant for stars with $T_{\rm eff}$ $\ge$ 3500 K. 
At the same time, $f_{\rm V}$ is in the range of 0.01--0.05 for both molecules. This indicates that cool starspots in M dwarfs may be relatively small structures (perhaps similar to magnetic pores on the Sun) with large surface coverage and highly intermittent polarity.

The contrasting behavior between the mol1 FeH and TiO $f_{\rm I}$ can be explained by differences in their formation heights. Figure~\ref{fig:cf} shows averages of the TiO and FeH line core contribution functions for the atmosphere models with $T_{\rm eff}=$ 3200\,K and 3800\,K. Contribution functions for other temperatures are not plotted, since Figure \ref{fig:tabmol}c shows that the differences in filling factors depend on the photosphere temperature.
 One can see that the formation height of FeH peaks deeper and that of TiO higher in the atmosphere for both temperatures. Contribution functions of the considered atomic line cores peak at intermediate heights. Hence, the TiO $f_{\rm I}$ probes a higher atmosphere where the magnetic fields fill about 0.7 of it in all M dwarfs, while the FeH $f_{\rm I}$ probes a lower atmosphere where the magnetic field filling factor increases from 0.3 in M1 to 0.7 in M5--M7 dwarfs.

\subsection{The magnetic field complexity}\label{complex}

We have found that in early M dwarfs (M1--M3) the cool spot area is smaller in the lower atmosphere (0.3) and larger in the higher atmosphere (0.8). This is consistent with magnetic spot models expanding from a lower to upper atmosphere \citep[e.g.,][]{rempeletal2009}. Using the model atmospheres, we have computed the so-called plasma $\beta$ parameter, which is the ratio of the gas pressure to the magnetic pressure. The magnetic field strength producing the pressure equal to the ambient gas pressure (i.e., when $\beta=1$) is sometimes called the equipartition magnetic field, $B_{\rm eq} $ (not to be confused with the definition arising from the equipartition theorem). For the average magnetic field of 5\,kG, which we have found in M1--M7 dwarfs, $\beta$ $<$ 1 throughout the entire atmosphere of all M dwarfs (above the optical depth 1 at 1.2\,$\mu$m). This indicates that plasma in M dwarf atmospheres is structured by the magnetic field. On the Sun, this is only observed in the solar chromosphere and corona as well as in sunspot umbra, while in the photosphere outside sunspots plasma motion is controlled by thermodynamic processes.

In late M dwarfs (M5--M7), the cool spot area is large (on average 0.7) in both the lower and higher atmosphere. This may indicate that the magnetic flux emergence is larger in these stars, and strong magnetic fields fill gaps in the lower atmosphere. Thus, the entire atmosphere of a late M dwarf is more complex and active. More frequent and more powerful flares observed in late M dwarfs \citep{yangetal2017, mondriketal2019} support this interpretation of our results.

A sketch of different magnetic regions in M dwarf atmospheres including spots expanding toward higher layers as well as small-scale, network magnetic fields in the lower atmospheres is given in Figure~\ref{fig:skizze}. It can be seen that atomic and molecular lines are formed at different optical depths which allows us to probe changing magnetic strengths and structure size at different heights. 

The complexity $\epsilon_{\rm IV}$  of the magnetic field quantity was found to vary with the height in the atmosphere. Also, the height-dependence itself depends on the stellar mass. The largest complexity is seen in network fields, with atomic  $f_{\rm I}=0.7-1.0$ (except for one star) and $f_{\rm V}=0.01-0.2$, leading to about 70--99\%\ of hidden network fields. Cool spots probed by molecular lines are less entangled than the network, implying that, locally, they are larger-scale structures than network fields, similar to the Sun. The complexity range in spots is about 10--70\%. Furthermore, we have found that the complexity  tends to increase toward lower mass M dwarfs.

In contrast, ZDI maps based on only atomic Stokes V \citep[e.g.,~][]{donati2013, morinetal2010} reveal a very small part of the stellar magnetic flux. Thus, interpretation of this residual flux as representative of the global magnetic field in terms of different components (toroidal and poloidal fields, and the degree of axisymmetry of the poloidal field) is biased due to the underestimated true complexity of the field. A discussion of tomographic field topology reconstructions and their limitations is given by \citet{berdyugina2009} and \citet{kochukhovetal2017}. Another definition of the complexity of  magnetic fields was used by \citet{shulyaketal2014}: as the minimum number of magnetic field components required to fit the observed Stokes I line profiles. These approaches, based only on one Stokes parameter, cannot fully describe the complexity of magnetic fields on M dwarfs (and other stars). As we have found, the complexity varies in three dimensions, depending on the height in the atmosphere.

\section{Summary}\label{conc}

Low-mass stars with fully convective interiors exhibit magnetic phenomena similar to those observed on the Sun, although they lack  an interface layer where dynamo processes are thought to take place. Thus, exploring magnetic fields and their topology and complexity provides a key tool for understanding stellar dynamos. In this work, we have determined temperatures of the photosphere and magnetic regions as well as mean magnetic field strengths and their filling factors for nine M dwarfs (M1--M7). We employed observed intensity and circular polarizaion spectra of selected atomic and molecular lines and a $\chi^2$ analysis based on model atmospheres and polarized radiative transfer.

We have found that all studied M dwarfs are highly magnetic with the average field of $\sim$5\,kG covering up to 100\%\ of the star, similar to the conclusions by \citet{krullvalenti2000}. In particular, two kinds of magnetic regions have been distinguished: one is significantly cooler than the photosphere (starspots) and has a filling factor of about 60--80\%\ in the upper atmosphere, while the other one is as hot as the photosphere (network) and has a filling factor 70--100\%\ in the lower atmosphere. By analyzing the filling factors determined from intensity and circular polarization spectra, we have determined the complexity of stellar magnetic fields as the difference between these two filling factors. It was found to increase toward lower-mass stars. 

In addition to "surface" inhomogeneities, we have also identified differences with the height in the atmosphere depending on the analyzed atomic or molecular species. In M dwarfs, FeH forms deeper and TiO higher in the atmosphere, while atomic lines form at intermediate heights. This has allowed us to establish that cool magnetic starspots in the lower atmosphere are smaller and less complex in earlier M dwarfs than in later M dwarfs, while in the upper atmosphere they are large and highly entangled in all M dwarfs. The network magnetic fields are more complex and entangled than starspots and fill practically the entire atmosphere of all M dwarfs.

More specifically, our findings are as follows:

The unspotted photosphere temperature ranges from 3800K for an M1 dwarf to 3100K for an M7 dwarf. These are somewhat higher than average effective temperatures (photosphere + cool spots) of M dwarfs determined from observed colors or spectral fitting without accounting for cool spots.

Two kinds of magnetic regions with clearly different temperatures exist in M dwarf atmospheres: cold starspots and warm network (similar to the Sun). Their temperatures vary with the spectral class of stars. The network temperature is the same (within uncertainties) as the unspotted photosphere.

The magnetic field strength is on average 5\,kG in both starspot and network features, and this does not vary across the M1--M7 spectral class range. We conclude that this is an implication that magnetic structures become locally smaller in scale toward lower mass M dwarfs. Also, the plasma $\beta$ parameter is smaller than 1 throughout the entire atmosphere of all M dwarfs (above the optical depth 1 at 1.2\,$\mu$m), indicating that the plasma in M dwarf atmospheres is structured by the magnetic field, like in the solar chromosphere, corona and sunspot umbra.

Network fields on M dwarfs are of mixed polarity and highly entangled (possibly small-scale, as on the Sun). They fill almost the entire lower and intermediate height M dwarf atmosphere. Polarimetric measurements alone may underestimate their magnetic flux (the complexity is about 70--99\%).

Cool starspots on M dwarfs are also highly intermittent in polarity and appear as intermediate-scale features, similar to pores on the Sun. They are larger in a higher M dwarf atmosphere, where they can fill up to 80\%\ surface area. Because of the mixed polarity, more than half of these features may remain undetected when only polarimetric data are analyzed (the complexity is about 10--70\%).

The magnetic field complexity on M dwarfs (cf., entanglement) varies with the height in their atmospheres and depends on the spectral class.  There is an indication that the magnetic flux emergence is larger in later M dwarfs (lower-mass), where highly complex, strong magnetic fields fill the atmosphere in all dimensions.

This work is a major step toward a 3D reconstruction of magnetic fields in M dwarfs and a comparison with realistic 3D MHD models of sunspots and starspots. Analysing all Stokes IQUV parameters of both atomic and molecular lines in the next step of such a study is necessary for fully understanding magnetic structures on cool stars.


\begin{acknowledgements}
This work was supported by the European Research Council (ERC) Advanced Grant HotMol (ERC-2011-AdG 291659). SB thanks the Kavli Institute for Theoretical Physics (KITP), Santa Barbara, CA, for support during the program "Better Stars, Better Planets" as well as program's organizers and participants for fruitful discussions about stellar magnetism. We thank the anonymous referee for the detailed review which helped to improve our manuscript. 
\end{acknowledgements}


\bibliographystyle{aa}
\bibliography{journals,nafram}

\end{document}